\documentclass[pre,twocolumn,aps,showpacs]{revtex4}
\usepackage{graphicx}

\begin{document}

\author{I.M. Sokolov}
\affiliation{%
Institut f\"{u}r Physik, Humboldt-Universit\"{a}t zu Berlin,
Invalidenstr. 110, D-10115 Berlin, Germany}
\title{On solutions of a class of non-Markovian Fokker-Planck equations}
\date{\today}

\begin{abstract}
We show that a formal solution of a rather general non-Markovian
Fokker-Planck equation can be represented in a form of an integral
decomposition and thus can be expressed through the solution of the
Markovian equation with the same Fokker-Planck operator. This allows us to
classify memory kernels into safe ones, for which the solution is always
a probability density, and dangerous ones, when this is not guaranteed. The
first situation describes random processes subordinated to a Wiener process,
while the second one typically corresponds to random processes showing a
strong ballistic component. In this case the non-Markovian Fokker-Planck
equation is only valid in a restricted range of parameters, initial and
boundary conditions.
\end{abstract}


\maketitle

Many physical phenomena related to relaxation in complex systems are
described by non-Markovian Fokker-Planck equations in a form 
\begin{equation}
\frac{\partial }{\partial t}P({\bf x},t)=\int K(t-t^{\prime }){\cal L}P({\bf %
x},t^{\prime })dt^{\prime },  \label{NMFPE}
\end{equation}
where $K(t)$ is a memory kernel and where ${\cal L}$ is a linear operator
acting on variable(s) ${\bf x}$. Such equations are often postulated on the
basis of linear-response considerations for different physical situations
and in several cases can be more or less rigorously derived based on a
microscopic description. In the symmetric case, the usual form of the
operator ${\cal L}$ reads: 
\begin{equation}
{\cal L}P({\bf x},t)=D\left( - \nabla \frac{{\bf f}({\bf x})}{k_{B}T}P({\bf x}%
,t)+\Delta P({\bf x},t)\right) ,  \label{Onedim}
\end{equation}
where $D$ is the diffusion coefficient (here supposed to be
coordinate-independent) and ${\bf f}(x)=-\nabla U({\bf x})$ is a potential
force. Depending on boundary conditions, the operator Eq.(\ref{Onedim}) may or
may not possess an equilibrium state $P({\bf x},t)=W({\bf x})$, which
corresponds to the solution of the equation $-(k_{B}T)^{-1}{\bf f}({\bf x})W(%
{\bf x})+\nabla W({\bf x})=0$, so that $W({\bf x})=\exp (-U({\bf x})/k_{B}T)$
is a Boltzmann distribution. In this case Eq.(\ref{Onedim}) can be rewritten
in the form 
\begin{equation}
{\cal L}P({\bf x},t)=D\sum_{\alpha ,\beta }\frac{\partial }{\partial
x_{\alpha }}W({\bf x)}\frac{\partial }{\partial x_{\beta }}\frac{P({\bf x},t)%
}{W({\bf x)}},  \label{Zwa}
\end{equation}
which is known to appear naturally when describing thermodynamics of complex
systems when reducing their behavior to a few relevant variables
(thermodynamical observables $x_{\alpha }$) as is done e.g. in the Zwanzig's
approach \cite{Zwanz1}. Compared to the general form of Ref. \cite{Zwanz1},
Eq.(\ref{NMFPE}) lacks the drift term; in some cases this general form can
be reduced to Eq.(\ref{NMFPE}), say by a Galilean transformation, see \cite
{MKS,MeK2}.

The Eq.(\ref{NMFPE}) with a $\delta $-functional memory kernel $K(t)=\delta
(t)$ corresponds to a usual Fokker-Planck equation (FPE) describing
Markovian processes. The solution of this equation is known to be a proper
probability density (so that $P({\bf x},t)\geq 0$ and $\int P({\bf x},t)d%
{\bf x}=1$) if the stationary state exists (i.e. whenever the Fokker-Planck
operator possesses a zero eigenvalue), otherwise it is a non-proper
probability density ($P({\bf x},t)\geq 0$ and $\int P({\bf x},t)d{\bf x}\leq
1$).

Many special forms of memory kernels are of interest. We note that
fractional Fokker-Planck equations widely discussed as a relevant
mathematical tool for the description of many complex phenomena \cite{MeK2}
belong just to the class described by Eq.(\ref{Zwa}) with $K(t)$ being a
power function of $t$: $K(t)\propto t^{-\alpha }$, and that the so-called
distributed-order fractional equations, introduced on the phenomenological
basis in Ref. \cite{Caputo} and describing slow processes lacking scaling 
\cite{Chech} correspond to related kernels in a form $K(t)\propto \int
f(\alpha )t^{-\alpha }d\alpha $. On the other hand, much less exotic
exponential kernels, describing the rather fast memory decay, are
ubiquitous. More complex kernels are encountered when describing reactions in
polymer systems \cite{Cherayil,Ghosh}.

The subdiffusive processes described by fractional Fokker-Planck equations
with $1<\alpha <2$ are known to be subordinated to a Wiener process \cite{Sok1,Barkai},
so that the solution of this equation can be obtained through
an integral transformation of the solution of a usual (Markovian) FPEs with
the same potential, initial and boundary conditions. As we proceed to show,
some analogue statements can be done also for the general version of the
non-Markovian FPE. The properties of such a transform and some important
consequences of its existence will be discussed in what follows.

Let us show that the formal solution of the non-Markovian Fokker-Planck
equation can be obtained in a form of an integral decomposition 
\begin{equation}
P({\bf x},t)=\int_{0}^{\infty }F({\bf x},\tau )T(\tau ,t)d\tau ,
\label{Subord}
\end{equation}
where $F({\bf x},\tau )$ is a solution of a Markovian FPE with the same
Fokker-Planck operator ${\cal L}$%
\begin{equation}
\frac{\partial }{\partial t}F({\bf x},t)={\cal L}F({\bf x},t),  \label{Mark}
\end{equation}
and for the same initial and boundary conditions, and the function $T(\tau
,t)$ is connected with the memory kernel $K(t)$ \cite{Sok1,Barkai}. Parallel
to Ref. \cite{Sok1} we shall call $\tau $ the internal variable of
decomposition, and ${\bf x}$ and $t$ its external variables. Moreover, we
show that the Laplace-transform $\tilde{T}(\tau ,u)$ of $T(\tau ,t)$ in its
external variable, $\tilde{T}(\tau ,u)=\int_{0}^{\infty }T(\tau ,t)e^{-ut}dt$
reads: 
\begin{equation}
\tilde{T}(\tau ,u)=\frac{1}{\tilde{K}(u)}\exp \left[ -\tau \frac{u}{\tilde{K}%
(u)}\right] ,  \label{Tlap}
\end{equation}
where $\tilde{K}(u)$ is a Laplace-transform of the memory kernel $K(t)$. Eq.(%
\ref{Tlap}) means that the Laplace-transform of $P({\bf x},t)$ in its
temporal variable reads: 
\begin{eqnarray}
\tilde{P}({\bf x},u) &=&\int_{0}^{\infty }dte^{-ut}\int_{0}^{\infty }d\tau F(%
{\bf x},\tau )T(\tau ,t) \nonumber \\ 
 &=&\int_{0}^{\infty }d\tau F({\bf x},\tau )\tilde{T}%
(\tau ,u) \nonumber   \\
&=&\int_{0}^{\infty }d\tau F({\bf x},\tau )\frac{1}{\tilde{K}(u)}\exp \left[
-\tau \frac{u}{\tilde{K}(u)}\right] \nonumber  \\ 
 &=&\frac{1}{\tilde{K}(u)}\tilde{F}\left(%
{\bf x},\frac{u}{\tilde{K}(u)}\right) , \label{Subor1} 
\end{eqnarray}
where $\tilde{F}({\bf x},u)$ is a Laplace-transform of $F({\bf x},\tau )$ in
its second variable $\tau $. Let us now note that the Laplace-transform of
the non-Markovian FPE, Eq.(\ref{NMFPE}) reads: 
\begin{equation}
u\tilde{P}({\bf x},u)-P(x,0)=K(u){\cal L}\tilde{P}({\bf x},u),  \label{Lapl1}
\end{equation}
where $P(x,0)$ is the initial condition. Inserting the form, Eq.(\ref{Subor1}%
), into Eq.(\ref{Lapl1}) one gets: 
\begin{equation}
\frac{u}{K(u)}\tilde{F}\left( {\bf x},\frac{u}{K(u)}\right) -P({\bf x},0)=%
{\cal L}\tilde{F}\left( {\bf x},\frac{u}{K(u)}\right) .  \label{Lapl2}
\end{equation}
Introducing a new variable $s=u/K(u)$ we rewrite Eq.(\ref{Lapl2}) in a form 
\begin{equation}
s\tilde{F}\left( {\bf x},s\right) -P({\bf x},0)={\cal L}\tilde{F}\left( {\bf %
x},s\right) ,  \label{Lapl3}
\end{equation}
in which one readily recognizes the Laplace-transform of an ordinary,
Markovian FPE, Eq.(\ref{Mark}), with the same initial condition $P({\bf x}%
,0) $. This completes our proof. Thus, the solution of a non-Markovian
Fokker-Planck equation of the type of Eq.(\ref{NMFPE}) in the Laplace domain
is connected with the solution of the regular Fokker-Planck equation through 
\begin{equation}
\tilde{P}({\bf x},u)=\frac{1}{\tilde{K}(u)}\tilde{F}\left( {\bf x},\frac{u}{%
\tilde{K}(u)}\right).  \label{ResFin}
\end{equation}
In the time domain this corresponds to Eq.(\ref{Subord}), where $T(\tau ,t)$ is given by Eq.(%
\ref{Tlap}).

The existence of the formal solution of the non-Markovian FPE in form of Eq.(%
\ref{ResFin}) brings several advantages: It gives an analytical tool to
express the solution of the non-Markovian problem through the solution of
the Markovian one, which is often known (at least for simple potentials and
simple boundary conditions). Even if the solutions are not known
analytically, a numerical procedure based on Eq.(\ref{Subord}) can be much
simpler than the direct solution of Eq.(\ref{NMFPE}). Moreover, in many
cases the non-negativity of the solution of a non-Markovian FPE (assumed to
be a probability density) can be easily proved {\it without solving the
equation}. This is true for a wide class of relaxation processes
subordinated to a Wiener process, i.e. for the equations with ''safe''
kernels ({\it vide infra}).

Note that $F({\bf x},t)$ is for each ${\bf x}$ a nonnegative function of $t,$ since it
is a (possibly, non-proper) probability density function (pdf) in ${\bf x}$. A
function $f(u)$ is a Laplace-transform of a nonnegative function defined on $%
\left[ 0,\infty \right) $ if and only if $f(u)$ is completely monotone, i.e. 
$f(0)>0$ and $(-1)^{n}f^{(n)}(u)\geq 0$, see  Chap. XIII of Ref. \cite{Feller}. 
Remember now, that $P({\bf x},u)=\tilde{F}\left( {\bf x},u/\tilde{K}(u)\right) /%
\tilde{K}(u)$ where the function $\tilde{F}\left( {\bf x},s\right) $ is
completely monotone in its second variable. This allows us to classify all
kernels into the ''safe'' ones, for which $\tilde{P}({\bf x},u)$ is completely
monotone for {\it any} completely monotone function $\tilde{F}\left( {\bf x}%
,u\right) $, and the ''dangerous'' ones, when this is not the case. Noting
that the product of two completely monotone functions is a completely
monotone function and that a function of the type $f(g(u))$ is completely monotone,
if $f(s)$ is
completely monotone and if the function $g(u)$ is positive and possesses a
completely monotone derivative \cite{Feller}, we can easily formulate a
sufficient condition for safety: It is the case if both functions, $\tilde{K}(u)$
and $u/\tilde{K}(u)$ are positive and possess completely monotone
derivatives. As we proceed to show, in this case $T(\tau ,t)$ is a pdf in
its first variable. The kernels for which this is the case are ''safe'' in
the sense that whatever the Fokker-Planck operator ${\cal L}$ is (i.e.
whatever the potential, the initial and the boundary conditions are), the
solutions of the non-Markovian FPE will be nonnegative and physically sound.
The dangerous kernels correspond to the situations when the physical
solutions of the non-Markovian Fokker-Planck equations exist only in the
restricted domain of parameters.

The function $T(\tau ,t)$ is always normalized to unity with respect to
variable $\tau $. To see this, let us consider $J(t)=$ $\int_{0}^{\infty
}d\tau T(\tau ,t)$. Its Laplace transform is $\tilde{J}(u)=\int_{0}^{\infty
}d\tau \tilde{T}(\tau ,u)=\tilde{K}^{-1}(u)\int_{0}^{\infty }d\tau \exp
\left[ -\tau u/\tilde{K}(u)\right] =1/u$, so that $\tilde{J}(t)\equiv 1$. On
the other hand, $T(\tau ,t)$ may or may not be a probability distribution of 
$\tau $ on $\left[ 0,\infty \right) $, depending on whether this function is
non-negative or may take negative values. For all safe kernels $T(\tau ,t)$ 
{\it is} a probability distribution: $\tilde{T}(\tau ,u)$ has just the form $%
\tilde{T}(\tau ,u)=\exp \left[ -\tau u/\tilde{K}(u)\right] /\tilde{K}(u)$,
i.e. corresponds exactly to the form mentioned above where we take $\exp
(-\tau u)$ instead of function $\tilde{F}$. The non-negativity of the
solutions of the non-Markovian FPEs then immediately follows from the fact
that the integrand in Eq.(\ref{Subord}) is a product of two non-negative
functions.

Let us now consider a few examples.

{\it Example 1}. As a simplest example let us consider the Markovian
situation, in which $K(t)=\delta (t)$, so that $K(u)=1$. The function $%
\tilde{T}(\tau ,u)=K(u)^{-1}\exp \left[ -\tau u/K(u)\right] =$ $\exp \left[
-\tau u\right] $, so that $T(\tau ,t)=\delta (t-\tau )$, and the
decomposition, Eq.(\ref{Subord}), is an identity transform.

{\it Example 2}. An example of a safe kernel is a power-law kernel $%
K(t)\simeq t^{-\alpha }$ with $1<\alpha <2$: both its Laplace-transform is $%
\tilde{K}(u)=\Gamma (1-\alpha )u^{\alpha -1}$ and the function $u/\tilde{K}%
(u)=u^{2-\alpha }/\Gamma (1-\alpha )$ are positive and have a completely
monotone derivative. Note that such power-law kernels just correspond to the
fractional Fokker-Planck equations (with the additional fractional
derivative of the order $\gamma =\alpha -1$ in their right-hand side, i.e.
with $0<\gamma <1$), which got now to be popular tools in describing slow
relaxation \cite{MeK2}. These equations are absolutely safe \cite
{Sok1,Barkai}. The same is valid for the kernels of the distributed-order
equations, $K(t)\propto \int f(\alpha )t^{-\alpha }d\alpha $ as long as
the function $f(\alpha )$ vanishes outside of the interval $(1,2)$, Ref. 
\cite{Chech}.

{\it Example 3}. As an example of a dangerous kernel we consider a simple
exponentially decaying one $K(t)=r\exp (-rt)$ (the form is taken to be
normalized in a way that for $r\rightarrow \infty $ it tends to a $\delta $%
-function). The Laplace-transform of this kernel reads: $\tilde{K}%
(u)=r/(u+r) $, so that $u/\tilde{K}(u)=\left( u^{2}+ru\right) /r$. Here the
first and the second derivative of the last function have the same positive
sign; thus it is not completely monotone. Let us show, that the
non-Markovian FPE with such a kernel may lead to negative solutions.

This really is the case if the system's behavior in a constant field is
considered. In what follows we restrict ourselves to a one-dimensional
situation. The Green's function solution of the FPE in a constant field
(initial condition $F(x,0)=\delta (x)$) reads: 
\begin{equation}
F(x,\tau )=\frac{1}{2\sqrt{\pi D\tau }}\exp \left[ -\frac{\left( x-\mu f\tau
\right) ^{2}}{4D\tau }\right]  \label{Gauss}
\end{equation}
with $\mu=D/k_BT$, so that its Laplace-transform in its temporal variable is: 
\begin{eqnarray}
\tilde{F}(x,s) &=&\frac{\exp (\mu fx/2D)}{2\sqrt{\pi D}} \times \nonumber  \\
&\times& \int_{0}^{\infty }\frac{1}{\sqrt{t}}\exp \left[ -\left( \frac{\mu ^{2}f^{2}}{4D}+s\right) t-%
\frac{x^{2}}{4D}t^{-1}\right] dt=  \nonumber \\
&=&\frac{\exp (2\zeta \lambda )}{2\sqrt{D}}\frac{1}{\sqrt{\zeta ^{2}+s}}\exp
\left[ -2\sqrt{(\zeta ^{2}+s)\lambda ^{2}}\right]
\end{eqnarray}
(see 2.3.16.2 of Ref.\cite{BP}), where the variables $\lambda =x/2\sqrt{D}$
and $\zeta =\mu f/2\sqrt{D}$ are introduced. The function $\tilde{P}(x,u)$
is obtained from $\tilde{F}(x,s)$ by multiplying by $1/\tilde{K}(u)$ and by
substitution $s=u/\tilde{K}(u)$, so that 
\begin{eqnarray}
\tilde{P}(x,u)&=&\frac{\exp (2\zeta \lambda )}{2\sqrt{D}}\frac{(u+r)/r}{\sqrt{%
\zeta ^{2}+u(u+r)/r}} \times  \nonumber \\
&\times& \exp \left[ -2\left| \lambda \right| \sqrt{\zeta
^{2}+u(u+r)/r}\right]  \label{Result}
\end{eqnarray}
This function is not completely monotone: Its first derivative (which has to
be always negative in the case of a pdf) changes sign, getting (for small $u$%
, i.e. in the long-time asymptotic) positive for $\zeta >(2r)^{-1}\left(
\left| \lambda \right| r^{2}+\sqrt{\lambda ^{2}r^{4}+2r^{3}}\right) $ (here
we took $\zeta >0$, so that the overall distribution moves to the right).
The oscillations occur initially at small $\left| \lambda \right| $,
corresponding to the initial position. Since the
maximum of the pdf moves to the right, they occur at the left flank of the
distribution, and for $\zeta >\zeta _{c}=\sqrt{r/2}$. Thus, if the force $f$
is strong enough, $f>\sqrt{2rD}/\mu $, the solution of non-Markovian FPE
ceases at long times to be a probability density, except for the
Markovian case $r\rightarrow \infty $. On the other hand, for the force-free
case of pure diffusion ($\zeta =0$) we have 
\begin{equation}
\tilde{P}(x,u)=\frac{1}{2\sqrt{D}}\frac{\sqrt{(u+r)/r}}{\sqrt{u}}\exp \left[
-2\left| \lambda \right| \sqrt{u\left( u+r\right) /r}\right] ,
\label{Danger!}
\end{equation}
which is a completely monotone function defining a pdf. For $u$ small ($t$
large) this function tends to a form corresponding to a Gaussian: 
\begin{equation}
\tilde{P}(x,u)\simeq \frac{1}{2\sqrt{D}}\frac{1}{\sqrt{u}}\exp \left[
-2\left| \lambda \right| \sqrt{u}\right] ,  \label{G1}
\end{equation}
which is our Eq.(\ref{Gauss}) with $f=0$, while for large $u$ (small $t$) we
have 
\begin{equation}
\tilde{P}(x,u\rightarrow 0)=\frac{1}{2\sqrt{Dr}}\exp \left[ -2\left| \lambda
\right| u/\sqrt{r}\right]  \label{Largeu}
\end{equation}
corresponding to 
\begin{equation}
P(x,t)\simeq \frac{1}{2\sqrt{Dr}}\delta \left( \frac{2\left| \lambda \right| 
}{\sqrt{r}}-t\right) =\frac{1}{2}\delta (\left| x\right| -\sqrt{Dr}t).
\end{equation}
At early times an initial pulse 
propagates as a wave, while at later times the propagation gets diffusive,
Ref. \cite{KMS}. 
Note that $Dr$ has a dimension of velocity squared (so that $D=v^{2}/r$
where $v$ is the typical velocity and $\tau _{c}=1/r$ is the correlation
time). Thus, at short times $P(x,t)=\frac{1}{2}\delta (\left| x\right| -vt)$%
, and the overall equation describes the transition from a ballistic to a
diffusive propagation, i.e. a kind of a {\it Drude model}. The mean-free
path in the model is exactly $v\tau _{c}=\sqrt{D/r}$. The breakdown of the
physical solution for larger forces gets now a clear physical meaning: The
case $f>\sqrt{2rD}/\mu $ corresponds to the situation when the mean velocity
gain on the mean free path is larger than the rms velocity $v=%
\sqrt{Dr}$, clearly the case in which the diffusion coefficient $D$ can no
more be considered as force-independent (which is only possible for $\mu
f\ll v$ where the force enters as a perturbation).

Thus, our analysis shows that the transition to non-positive solutions
denotes leaving the region of physical validity of the model: the fact that
the kernel is ''dangerous'' shows, that corresponding equations are only
reasonable in a restricted domain of parameters, initial and boundary
conditions, and that other conditions are unphysical.

The behavior of $P(x,t)$ for the exponential kernel and for $f=0$ is shown
in Fig.1, where the results of numerical inversion of Eq.(\ref{Result}) are
shown for $t=0.5$, 1, 2 and 3. Here only the part for $x>0$ is shown since $%
P(x,t)$ is an even function of $x$. The overall form of the distribution with the two side
peaks is typical for systems showing random-walk behavior with strong
ballistic component, like L\'{e}vy walks. At difference with the
L\'{e}vy-walk situation, the overall weight of the peaks decays very fast.
Thus, for $t=0.5$ they absorb more than one-half of the overall probability,
while for $t=3$ the most of the probability lies in the central part of the
distribution, whose form slowly tends to a Gaussian. 

\begin{figure}
\scalebox{0.35}{\includegraphics{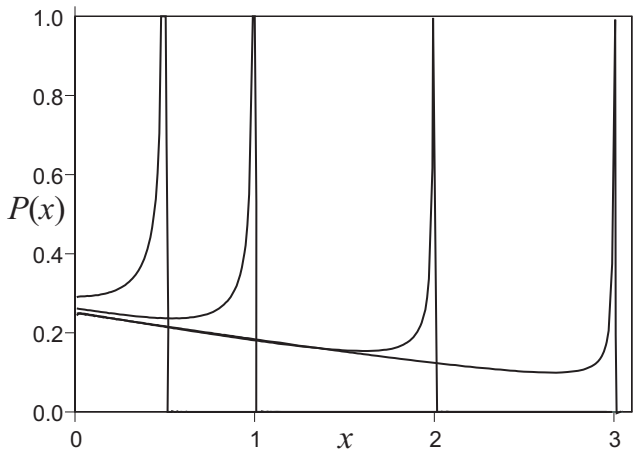}}
\caption{The time evolution of the Green's function solution of
a non-Markovian diffusion equation ($f=0$), see text for details. The
parameters are $D=$ $r=1$, so that the peaks move with the velocity $v=1$.
The curves correspond to $t=0.5$, 1, 2 and 3 (from left to right).}
\end{figure}

Note that the situation when the short-time behavior is ballistic and
corresponds to the distribution with pronounced side peaks (stemming
essentially from the solution of the Liouville equation) is typical (as a
short-time behavior) for all kernels $K(t)\,$which tend to a constant value
at $t\rightarrow 0$: For all of them $\tilde{K}(u)\simeq K(0)/u$ for $u$
large, so that Eq.(\ref{Largeu}) asymptotically holds. Turning to kernels
behaving at short times as a power-law, $K(t)\simeq t^{-\alpha }$
(corresponding to $K(u)\simeq \Gamma (1-\alpha )u^{\alpha -1}$), we note
that the kernels with $0<\alpha <1$ lead to the similar kind of behavior
(bimodal pdf), see Ref. \cite{MeK1}, where the peaks are the less pronounced
the larger is $\alpha $. On the other hand, safe power-law kernels with $%
1<\alpha <2$ lead to pdf's showing a single peak at zero. It is also
interesting to discuss the two other situations: The kernels starting from
zero and the strongly decaying power-laws. Now, a situation of a kernel
starting at zero, i.e. as $K(t)\simeq t^{\beta }$ with $\beta >0$,
corresponds to $\tilde{K}(u)\simeq u^{-1-\beta }$ and therefore to 
\[
\tilde{P}(x,u)=\frac{u^{\beta /2}}{2\sqrt{D}}\exp \left[ -2\left| \lambda
\right| u^{1+\beta /2}\right] 
\]
which is not a completely monotone function (its first derivative changes
sign at $u=\left[ 2\left| \lambda \right| (2+\beta )/\beta \right]
^{-2/(2+\beta )}$), and thus is not a Laplace-transform of a pdf. The same
is the case for the kernels with stronger divergence, $K(t)\simeq t^{-\alpha
}$ with $\alpha >2$. Here 
\[
\tilde{P}(x,u)=\frac{u^{-\alpha /2}}{2\sqrt{D}}\exp \left[ -2\left| \lambda
\right| u^{1-\alpha /2}\right] 
\]
is again not a completely monotone function: its second derivative $\tilde{P}%
_{uu}(x,u)$ (which is essentially a quadratic form in $\left| \lambda
\right| $) possesses a positive root for all $\alpha >2$. Thus, the
set of kernels which correspond to physical behavior (in a force-free case)
consists of kernels which behave at small $t$ as $K(t)\simeq t^{-\alpha }$
with $0\leq \alpha \leq 2$. All other kernels can be considered as
approximations which are not valid at short times.

Let us make some notes about the long-time asymptotic behavior. For $t$
large all integrable kernels correspond to the behavior $K(u)\rightarrow I$
with $I=\int_{0}^{\infty }K(t)dt$, and thus lead to 
\[
\tilde{P}(x,u)=\frac{1}{2\sqrt{D}}\frac{1}{\sqrt{Iu}}\exp \left[ -2\left|
\lambda \right| \sqrt{u/I}\right] 
\]
i.e. to the Gaussian behavior. All these kernels correspond essentially to
the processes which can at longer times be approximated by a Markovian
process. The kernels whose integral diverges are exemplified by the safe
power-law-like kernels $K(t)\propto t^{-\alpha }$ with $1\leq \alpha \leq 2$%
, where the divergence stems from the short-time behavior, and the dangerous
power-laws ($0\leq \alpha <1$) where the integral diverges at infinity. Both
of them correspond to non-decoupling memory. The situations are considered
in detail in Refs. \cite{Sok1,MeK2}. The growing kernels are definitely
unphysical.

Whenever the kernel is safe, the variable $\tau $ can be interpreted as an
operational time, and $T(\tau ,t)$ is a pdf of the operational time $\tau $
at physical time $t$, and our integral decomposition corresponds to a {\em %
subordination}. Whenever $T(\tau ,t)$ can be considered as pdf resulting
from a random process with nonnegative increments, we have to do with a
continuous-time random walk situation (CTRW) or its continuous limit. The
corresponding solutions of the non-Markovian FPEs (including the Green's
function solutions) can then be represented as the solutions of the ordinary
FPE corresponding to different final operational times; these solutions are
weighted with the distribution of this final operational time, which is given by the
pdf $T(\tau ,t)$. Thus, the ensemble of the sample paths corresponding to
a random process described by the non-Markovian FPE with a safe kernel can
be visualized as an ensemble of paths (random walks) of a process described
by a corresponding Markovian equation, taken not at a given time $t$, but
having different temporal ''lengths'' (duration).

The dangerous kernels correspond to the situation when some of these paths
enter with negative weight, so that the overall positiveness of the solution
can not in general be guaranteed. We note that the case of the 
exponential kernel (for which the non-Markovian FPE can be rewritten in 
the form of the telegrapher's equation) can be considered as
an approximation for a CTRW
with the waiting-time distribution being a difference of two exponentials
\cite {KMS}. However, neglecting higher terms in such an approximation
leads to the fact that the exponential kernel is dangerous, and that
the positiveness of the solution is not always guaranteed.

Let us now summarize our findings. We considered a formal solution of a
rather general form of a non-Markovian Fokker-Planck equation and have shown
that this can be represented in a form of integral decomposition. This
allows us to classify the memory kernels into safe ones, for which the
solution of the non-Markovian FPE is always a probability density, and
dangerous ones, when this is not guaranteed. In this case the non-Markovian
FPE is only valid in a restricted range of parameters, or under special
initial and boundary conditions. The examples of the non-Markovian FPE with
dangerous kernels considered render clear that such equations describe the
processes with strong ballistic component.

The author is grateful to Yossi Klafter for valuable discussions and to 
the Fonds der Chemischen Industrie for the partial financial support.

\end{document}